\documentclass{amsart}

\theoremstyle{definition}

\theoremstyle{remark}

\numberwithin{equation}{section}

\begin{document}

\title{SYMMETRIES AND LIE ALGEBRA OF RAMANUJAN EQUATION}


\author{Amlan K Halder}
\address{Department of Mathematics, Pondicherry University, Puducherry - 605014, India}
\curraddr{}
\email{amlanhalder1@gmail.com}
\thanks{}

\author{Rajeswari Seshadri}
\address{Department of Mathematics, Pondicherry University,
Puducherry - 605014, India}
\email{seshadrirajeswari@gmail.com}
\thanks{}
\author{R ~Sinuvasan}
\address{Department of Mathematics,VIT-AP, Amaravati-522237, Andhra Pradesh, India}
\email{rsinuvasan@gmail.com}
\thanks{}
\author{PGL Leach}
\address{School of Mathematics, Statistics and Computer Science, University
of KwaZulu-Natal, Durban, South Africa and}
\address{Institute for Systems Science, Durban University of Technology,
Durban, South Africa}
\email{leachp@ukzn.ac.za}
\thanks{}

\subjclass[2010]{34A05; 34A34; 34C14; 22E60.}

\keywords{symmetries; integrability; increase of order.}

\date{}

\dedicatory{}

\begin{abstract}
Symmetry analysis of Ramanujan's system of differential equations is performed by representing it as a third-order equation. A new system consisting of a second-order and a first-order equation is derived from Ramanujan's system. The Lie algebra of the new system is equivalent to the algebra of the third-order equation. This forms the basis of our intuition that for a system of first-order odes its infinite-dimensional algebra of symmetries contains a subalgebra which is a representation of the Lie algebra for any system or differential equation which can be obtained from the original system, even though the transformations are not point. 
\end{abstract}

\maketitle

\vspace{1.5cc}

\section{Introduction}
Ramanujan\cite{Ram 25} introduced three functions, P(q), Q(q) and R(q), defined for $\mid q\mid<1$,  and are noted as the Eisenstein series \cite{Ram 04, Ram 05}. They are represented as follows:

\begin{eqnarray}\label{1.1}
P(q) & = & 1-24\sum_{n=1}^\infty \frac{nq^n}{1-q^n},\nonumber \\
Q(q) & = & 1+240\sum_{n=1}^\infty \frac{n^3q^n}{1-q^n},\\
R(q) & = & 1-504\sum_{n=1}^\infty \frac{n^5q^n}{1-q^n}, \nonumber
\end{eqnarray}
These functions satisfy the following system of differential equations: 

\begin{eqnarray}\label{1.2}
q \frac{dP}{dq} &=& \frac{1}{12}(P^2-Q),\nonumber \\
q \frac{dQ}{dq} &=& \frac{1}{3}(PQ-R),\\
q \frac{dR}{dq} &=& \frac{1}{2}(PR-Q^2), \nonumber
\end{eqnarray}
and are called Ramanujan's differential equations. All of the results in (\ref{1.2}) were mentioned by Ramanujan in his second notebook. The functions $P, Q$ and $R$ are called classical Eisenstein series, $E_{2k}$ for $k=1,2,3,$  respectively. The three Eisenstein series and their differential equations are of enormous importance in many areas of Number Theory, for example, in Modular Forms. In \cite{Ram 16} it has been shown that, the solution of the system can be written as the Hypergeometric function with respect to its reduction to Riccati's equation using a particular one-parameter stretching group of transformations. This system of differential equations can be written as Chazy's equation under an Euler Transformation \cite{Ram 01, Ram 02, Ram 06, Ram 11, Ram 12}.\\

The objective of this article is to deduce different results for equation (\ref{1.2}) using only Lie group analysis \cite{Ram 03, Ram 21, Ram 22, Ram 26}. Our work deals with using the Lie Symmetry analysis to find the point symmetries of the third-order equation which can be obtained from the system (\ref{1.2}) and using them to reduce the third-order equation to Abel's equations of the First and Second Kinds. Symmetries are obtained using the mathematical software Maple (Maple is a trademark of Waterloo Maple Inc.)  and Mathematica \cite{Ram 03, Ram 08, Ram 09, Ram 10}. Implicit Hypergeometric solution of each of the Abel's equations of the First and Second Kind can be obtained. Furthermore, we study the system (\ref{1.2}) by writing it as a system of a second-order and a first-order differential equations. The Lie algebra obtained for this system is similar to that of  the third-order equation, i.e. a representation of  $A_{3,8}$ (We use the Mubarakzyanov Classification Scheme \cite{Ram 17, Ram 18, Ram 19, Ram 20}). The system (\ref{1.2}) has a representation of the Lie algebra $A_{3,8}$. Hence contained in the infinite symmetries of (\ref{1.2}) there is a representation of  $A_{3,8}$. \\

This forms the basis of our intuition that for a system of first-order odes, its infinite-dimensional algebra of symmetries contains a subalgebra which is a representation of the Lie algebra for any system or any differential equation which can be obtained from the original system of first-order ordinary differential equations (odes) even though the transformation is not point. Our intuition is supported by another system of differential equations, which is an analogue of Ramanujan's equations, derived by Ramamani. This system also satisfies our intuition. We consider the analogues of Ramanujan's differential equations for the classical Eisenstein series which were studied by V. Ramamani in the 1970's \cite{Ram 23, Ram 24}. The Singularity Analysis of the third-order equation obtained from (\ref{1.2}) yields positive results, whereas it fails for the new system derived from (1.2) and also for (1.2) itself. Recently, in of some of our other works \cite{Ram 13, Ram 14, Ram 15}, we have specified an alternate method to study a system of odes. \\

The paper is organized as follows: In Section $2,$ the symmetry analysis of the third-order equation is discussed. In Section $3,$ the Lie point symmetries of the new $2 + 1$ system are found and comprise a representation of the algebra $A_{3,8}$   ($A_{3,8}$ is also known as $sl(2, R)$). Along with that a set of symmetries from the infinite set of symmetries, which is a representation of  $A_{3,8}$   for (\ref{1.2}) has been shown. The symmetry analysis of the Ramamani equation and its Lie algebra, along with the system derived from it, is discussed herewith. In Section $4,$ we make mention of the singularity analysis test and the conclusion is provided with proper references.\\

\section{Symmetry Analysis}

We rewrite system (\ref{1.2}) as single nonlinear third-order differential equation

\begin{equation}\label{2.1}
2q^2P'''(q)=-6 q P''(q)+2 q P(q) P''(q)-3 q P'(q)^2+2 P(q) P'(q)-2 P'(q).
\end{equation}
The Lie Point Symmetries of (\ref{2.1}) are
\begin{eqnarray}\label{2.2}
\Gamma_1 & = & q\partial _q, \nonumber \\
\Gamma_2 & = &-\left(q \log \mid q\mid\right) \partial_q+ P \partial_P, \\
\Gamma_3 & = &-\left(q \log \mid q\mid^2\right)\partial _q+\left(2 P \log \mid q\mid+12\right)\partial_P,\nonumber
\end{eqnarray}
 which have an algebraic structure isomorphic to $A_{3,8}$. We consider $\Gamma_1$ for reduction. The canonical coordinates are $r=P$ and $s=\log \mid q\mid$. Then
 $$ v(r)=\frac{ds}{dr}=\frac{1}{q P'(q)}.$$
 Using these canonical coordinates we reduce (\ref{2.1}) to
\begin{equation}\label{2.3}
2 v''(r)=3 v(r)^2+2 r v(r) v'(r) +\frac{6 v'(r)^2}{v(r)}.
\end{equation}

Equation (\ref{2.3}) has the symmetry $\Gamma = r \partial_r-2 v \partial_v$. The canonical coordinates are $$ z=v(r) r^2 \quad\mbox{\rm and}\quad u(z)=\frac{1}{r^2(r v'(r)+2 v(r))},$$ whereby equation (\ref{2.3}) is reduced to

\begin{equation}\label{2.4}
u'(z)= \frac{(z^2-12z)}{2}u(z)^3+\frac{(-2z+14)}{2}u(z)^2-\frac{3u(z)}{z}.
\end{equation}
This is an Abel's equation of the first kind.
When we use the differential invariants
$$z=v(r)r^2 \quad\mbox{\rm and}\quad u(z)=v'(r)r^3,$$
equation (\ref{2.3}) is reduced to
\begin{equation}\label{2.5}
2(u(z)+2z) u'(z) = \frac{6 u(z)^2}{z}+\left(2 z+6\right)u(z)+3 z^2.
\end{equation}
This is Abel's equation of the second kind. The Abel's equations (\ref{2.4}) and  (\ref{2.5}) are not in any of the particular forms mentioned in \cite{Ram 27}. The solutions of Abel's equation of the First Kind and the Second Kind are given in terms of Implicit Hypergeometric Functions \cite{Ram 07}.

If we use $ \Gamma_2$ for reduction, equation (\ref{2.1}) is reduced to a second-order equation devoid of  point symmetries. $\Gamma_3$  and $\Gamma_1$ are equivalent, so the reductions are similar and equation (\ref{2.1}) reduced  to one of Abel's equations of the first or second kind.

\section{Lie Algebra of System of $ 2 + 1 $ equations}

The System of $ 2+1 $ equations comprises a second-order equation and a first-order equation obtained from (\ref{1.2}). After an Eulerian transformation,$$ P(q) \quad\mbox{\rm to}\quad P(e^q) \quad\mbox{\rm and}\quad Q(q) \quad\mbox{\rm to}\quad Q(e^q),$$ the newly derived autonomous system is,
\begin{eqnarray}\label{3.1}
-\frac{1}{12}P^2 +\frac{Q}{12}+P' &=&0,\nonumber \\
-\frac{1}{2}P^2 Q+\frac{Q^2}{2}+QP'+\frac{5}{2}PQ'-3Q'' &=&0.\\\nonumber
\end{eqnarray}

The Lie Point symmetries of (\ref{3.1}) are
\begin{eqnarray}\label{3.2}
\Gamma_{11} & = & \partial _q, \nonumber \\
\Gamma_{12} & = &q\partial_q-P\partial_P-2Q\partial_Q, \\
\Gamma_{13} & = &q^2\partial_q-(12+2Pq)\partial_P-4Qq\partial _Q,\nonumber
\end{eqnarray}
which have an algebraic structure isomorphic to $A_{3,8}$.

From (\ref{3.2}) the Lie Point symmetries of (\ref{1.2}) are
\begin{eqnarray}\label{3.3}
\Gamma_{21} & = &-2\partial _q \nonumber \\
\Gamma_{22} & = &-2q\partial_q+2P\partial_P+4Q\partial_Q+6R\partial_R \\
\Gamma_{23} & = &-q^2\partial_q+(12+2Pq)\partial_P+4Qq\partial _Q+6Rq\partial_R\nonumber
\end{eqnarray}
which constitute an algebraic structure isomorphic to $A_{3,8}$. As we see, for the system of first-order odes we have a representation isomorphic to, $A_{3,8}$ in the infinite-dimensional algebra of its symmetries.\\

Hence the Lie algebra of the third-order equation, system of $2+1$ equations and a subalgebra of (\ref{1.2}) is the same. This is our intuition that the Lie algebra of an ode derived from a system is the same as the Lie algebra or at least a subalgebra  of any new system derived from the original system and of the original system. Even though the transformation is not point.\\

We derive the Lie algebra of an analogue of Ramanujan's equation invented by Ramamani.
An analogue of Ramanujan's system was developed by Ramamani\cite{Ram 23,Ram 24}, using functions which were defined for $ \mid{q}\mid< 1 $ and equivalent to (\ref{1.1}),
\begin{eqnarray}\label{3.4}
{\it P}(q) & = & 1-8\sum_{n=1}^\infty \frac{(-1)^nnq^n}{1-q^n},\nonumber \\
{\it T}(q) & = & 1+24\sum_{n=1}^\infty \frac{nq^n}{1+q^n},\\
{\it Q}(q) & = & 1+16\sum_{n=1}^\infty \frac{(-1)^nn^3q^n}{1-q^n}, \nonumber
\end{eqnarray}
The functions $${\it P}(q),{\it T}(q)\quad\mbox{\rm and}\quad {\it Q}(q) $$ can be represented using Eisenstein series. They satisfy the differential equations

\begin{eqnarray}\label{3.5}
q \frac{d {\it P}}{dq} &=& \frac{1}{4}({\it P}^2-{\it T}),\nonumber \\
q \frac{d{\it T}}{dq} &=& \frac{1}{2}({\it PT}-{\it Q}),\\
q \frac{d{\it Q}}{dq} &=&({\it PQ}-{\it TQ}).\nonumber
\end{eqnarray}
\begin{eqnarray}\label{3.6}
q \left[(8 - 8 P(q) + 3 P(q)^2)P'(q) -8 q P'(q)^2 
+8 q (3 - P(q))P''(q)-q P'''(q)\right]+ \nonumber \\
\left[P(q)^2-P(q)
-4 q P'(q)\right] \left[P(q)^3 -8 q P(q)P'(q) +8 q (P'(q) + q P''(q))\right]=0.\nonumber\\
\end{eqnarray}

The sole Lie Point Symmetry of (\ref{3.6}) is
\begin{equation}\label{3.7}
\Gamma_1 =  q\partial _q. \nonumber \\
\end{equation}
The Lie algebra is a representation of $A_1$.
After an Eulerian transformation the new derived autonomous system  is
\begin{eqnarray}\label{3.8}
-\frac{1}{4}P^2 +\frac{Q}{4}+P' &=&0,\nonumber \\
-P^2Q+Q^2P+QP'+3PQ'-2QQ'-2Q''&=&0.\\\nonumber
\end{eqnarray}
We refer to (\ref{3.8}) as a $2+1$ system.
\begin{equation}\label{3.9}
\Gamma_1  =  \partial _q .\nonumber \\
\end{equation}
The Lie algebra is a representation of $A_1$.
Also the Lie algebra of (\ref{3.5}) contains $A_1$.

\section{Singularity analysis of system of equations}

The singularity analysis of (\ref{2.1}) can be directly related to that of the Chazy equation. The singularity analysis fails for the system of $2+1$ equations (\ref{3.1}) and for the system (\ref{1.2}).
\section{Conclusion}
We studied  Ramanujan's system of equations by means of symmetry analysis. Abel's equation of the first kind and of the second kind of first-order gives a solution in terms of  the Implicit Hypergeometric Function. The newly derived system, (\ref{3.1}), from the Ramanujan's system has the same Lie algebra as for (\ref{2.1}). The infinite symmetries of (\ref{1.2}) contain a representation of the Lie algebra, $A_{3,8}$, as shown in (\ref{3.3}). This forms the basis of our intuition. Our future works continue to study the reduction techniques for the newly derived system (\ref{3.1}) consisting of a second-order and a first-order differential equation for which we have obtained Lie point symmetries. Along with that we expect to derive the invariant solutions of Ramanujan's system.

\section{Acknowledgements}
AKH is grateful to NBHM Post-Doctoral Fellowship, Department of Atomic Energy 
 (DAE), Government of India, Award No: 0204/3/2021/R\&D-II/7242 for financial support.
PGLL acknowledges the support of the National Research Foundation of South Africa, the  
University of KwaZulu-Natal and the Durban University of Technology.

\end{document}